# Stable three-dimensional lattice solitons in spin-orbit-coupled Bose-Einstein condensates

**Liangwei Zeng,[1] Boris A. Malomed,[2] Yaroslav V. Kartashov,[3] and Xing Zhu[1,*]**

[1]*School of Arts and Sciences, Guangzhou Maritime University, Guangzhou 510725, China*

[2]*Instituto de Alta Investigación, Universidad de Tarapacá, Casilla 7D, Arica, Chile*

[3]*Institute of Spectroscopy, Russian Academy of Sciences, Fizicheskaya str. 5, 108840, Troitsk, Moscow, Russia*

**Corresponding author: xingzhu@gzmtu.edu.cn*

We address three-dimensional (3D) solitons maintained by spin-orbit coupling (SOC) in the binary self-interacting Bose-Einstein condensate (BEC) held in the 3D optical lattice (OL). The analysis reveals that the SOC-OL interplay results in the formation of stable full-vortex (FV) solitons, built as sets of four density peaks residing in neighboring wells of the lattice potential, with the superimposed global vortical phase, and site-centered semi-vortices (SV), in which the vorticity is present in only one component. The full-vortex solitons, with their specific phase textures, do not exist in a uniform BEC with SOC. Full-vortex and semi-vortex states in the binary self-attractive BEC are stable despite the possibility of the supercritical collapse in the 3D system. Such states also exist, as gap solitons, in the self-repulsive 3D system. In terms of the chemical potential and number of particles, the stability regions of the 3D full-vortex solitons and semi-vortices expand with the increase of the SOC strength and OL depth. The results open the route to the creation of 3D complexes of vorticity-carrying condensates that can be realized with existing experimental techniques.

Realization of stable multidimensional self-sustained states in nonlinear media is a problem of fundamental significance in diverse areas of physics, ranging from hydrodynamics, plasmas, acoustics and photonics to plasmonics and Bose-Einstein condensates [1-9]. A challenging problem is that, in the majority of these settings, the ubiquitous cubic nonlinearity, which is most frequently involved in the formation of self-sustained states, secures their stability only in one dimension (1D), while 2D and 3D solitons in uniform cubic nonlinear media are unstable due to the occurrence of the critical or supercritical collapse, respectively [9-13]. While unstable soliton-like states and their collapse have been observed in diverse optical [1,14] and matter-wave [15-18] setups, the search for stabilization mechanisms of soliton states, especially 3D ones, is a topic of great interest in fundamental and applied studies [19-23]. The stabilization of multidimensional states with embedded vorticity is even more challenging problem, as, in addition to the collapse, they are subject to stronger azimuthal instabilities that tend to split them into fragments [24-30]. A powerful mechanism allowing the stabilization of multidimensional solitons is the use of trapping potentials, including harmonic-oscillator [31-33] and periodic ones [34]. These potentials support not only stable 1D solitons [35], but may also stabilize 2D [36-40] and 3D [41-45] ones. Despite the prediction of this possibility for the stabilization of 3D solitons, it has not yet been experimentally demonstrated in self-attractive BEC, while in photonics they were observed only as transient states [46,47]. Another well-known stabilization mechanism for multidimensional self-sustained states relies on the competition of different nonlinearities, as it occurs for the recently observed quantum droplets in BEC, whose stability is secured by the self-repulsive quantum-mechanical correction to the effectively attractive mean-field energy, which is a residue of the nearly exact balance of the inter- and intraspecies interactions in binary BEC with opposite signs [5-8,48-52]. It was also predicted that the same mechanism can maintain stable 3D quantum droplets with intrinsic vorticity [53]. Nonlocality [54,55] or inhomogeneity [56,57] of the nonlinear response can also give rise to stable self-trapped multidimensional states.

Realization of spin-orbit coupling (SOC) in Bose-Einstein condensates [58-60] in different dimensions has opened broad prospects for the investigation of the impact of artificial gauge fields on the evolution of condensates in a great variety of settings [61-74]. SOC not only qualitatively changes the linear spectrum of the spinor system, but also results in the appearance of new self-sustained nonlinear states. As a result, even in the 1D case exotic stripe phases may appear, solitons forming in SOC-BEC do not obey the Galilean invariance [75-80], and moving solitons collide inelastically. An important effect of SOC is the stabilization of nonlinear states. Thus, 2D SOC prevents the critical collapse under the action of the cubic self-attraction and supports stable 2D solitons of the *semi-vortex* (SV) type, by pushing their norm below the collapse threshold [81-84], although it does not produce stable states carrying vorticity in both components. In Ref. [85] it was found that an inhomogeneous, rather than uniform, SOC can sustain 2D linear localized zero-vorticity and vortical modes, and even provide confinement in the presence of repulsive interactions. Further, 3D SOC enables the formation of metastable SV solitons in 3D self-attractive uniform media [86], but it does not yield stability of full vortex solitons carrying vorticity in both components of the spinor wavefunction. In this respect, the combined effect of SOC and an optical-lattice (OL) potential, that can be implemented by means of available experimental techniques, can be promising for the realization of new types of stable multidimensional solitons. Lattice solitons in SOC-BEC setups have been studied with 1D [87-96] and 2D [97-103] periodic potentials, but the most promising case, based on the interplay between the nonlinearity, SOC and 3D OL potentials, has never been considered before.

This work reports that this general setting allows the creation of both stable semi-vortex (SV) solitons and of novel 3D stable full-vortex (FV) states carrying vorticity in both components. We find them not only for attractive interactions in the semi-infinite spectral gap of the lattice potential, but also as *gap solitons* [104,105] for repulsive interactions, in finite spectral gaps. The OL, even if it is shallow, strongly expands the stability domain of the SV solitons. The stability domain for FV solitons rapidly expands with the increase of the SOC strength. The cooperation of the two factors, SOC and OL potential, each of which acts toward improving the stability of self-trapped states in 3D attractive media, allows one to produce stable soliton states with new symmetries and spin textures. Demonstrated stability of FV solitons in 3D system is one of our central results, since in general it is very hard to make such states stable even in 2D geometries with attractive nonlinearity (the reason behind this is that vortical phase in individual elements forming vortex complex makes the interactions of neighboring elements very complicated). Therefore, one of our central results is not mere prediction of a new type of solution, but illustration that the combination of the above mentioned mechanisms can stabilize solitons with complex internal structure in 3D geometries with attraction – the situation that is very hard to achieve in physics. The theoretical results reported in the paper suggest the possibilities for the creation of such nontrivial vortex states in experiments with ultracold bosonic gases.

The evolution of the mean-field wavefunction $\mathbf{\Psi}(\mathbf{r},t)=(\Psi_1,\Psi_2)^{\mathrm{T}}$ of the binary condensate under the action of the 3D OL potential is governed by the scaled Gross-Pitaevskii equation:

$$i\frac{\partial\mathbf{\Psi}}{\partial t}=-\frac{1}{2}\nabla^2\mathbf{\Psi}-i\lambda(\nabla\cdot\boldsymbol{\sigma})\mathbf{\Psi}+\mathcal{U}(\mathbf{r})\mathbf{\Psi}+g(\mathbf{\Psi}^\dagger\mathbf{\Psi})\mathbf{\Psi}, \quad (1)$$

where $\mathbf{r}=(x,y,z)$ is the set of spatial coordinates, $\lambda$ is the SOC strength, and $\boldsymbol{\sigma}=(\sigma_{\mathrm{x}},\sigma_{\mathrm{y}},\sigma_{\mathrm{z}})$ is the vector of the Pauli matrices [60], while $\mathcal{U}(\mathbf{r})=-\mathcal{U}_0[\cos(2x)+\cos(2y)+\cos(2z)]$ is the OL potential with depth $\mathcal{U}_0$ and equal periods $d_{x,y,z}=\pi$ in all three directions. We consider the nonlinearity with equal strengths of the self- and cross-interactions that may be attractive $(g=-1)$ or repulsive $(g=+1)$, as it is the physically relevant situation for binary BEC. An estimate of the corresponding physical parameters is provided, in particular, by those reported in Ref. [59]. In that case, about $1.5\times10^5$ $^{87}$Rb atoms (a sufficiently large number to create clearly visible 3D states) were confined by the optical dipole trap with frequencies $(f_{\mathrm{x}},f_{\mathrm{y}},f_{\mathrm{z}})=2\pi\times(45,45,55)\ \mathrm{Hz}$. The recoil energy was $E_r=\hbar^2k_0^2/2m$, where $m$ is the atomic mass and $k_0$ is the wavenumber of the lattice-inducing lasers (the wavelength was $767\ \mathrm{nm}$).

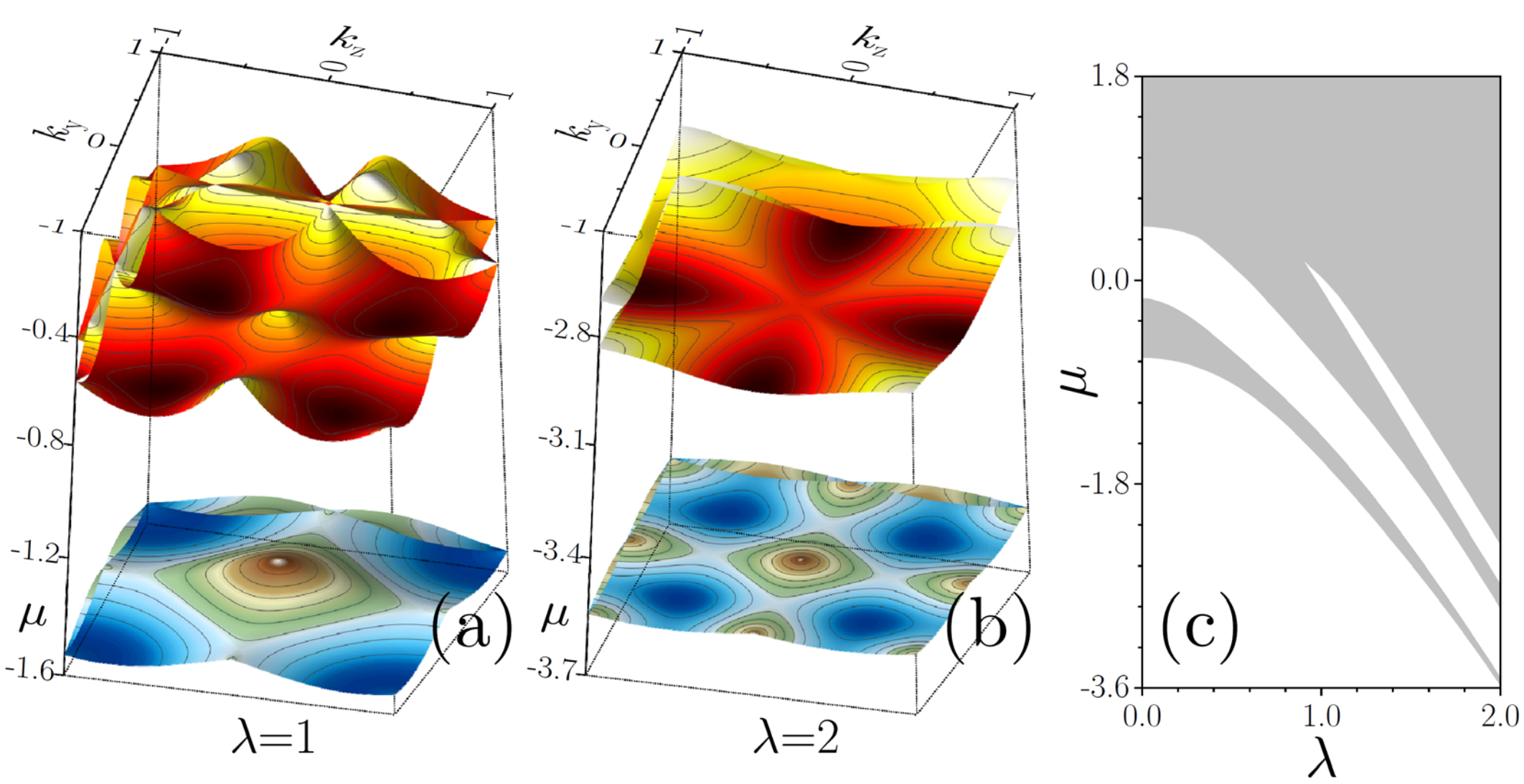

Fig. 1. Panels (a) and (b) show dependences of eigenvalues of Bloch waves $\mu_\nu(\mathbf{k})$ (the vertical axis) in potential $\mathcal{U}(\mathbf{r})$, belonging to six lowest bands (i.e. $\nu=1,\dots,6$), versus $k_{\mathrm{y}}$ and $k_{\mathrm{z}}$ components of the Bloch momentum, at fixed $k_{\mathrm{x}}=0$, for the SOC strength $\lambda=1$ (a) and $2$ (b). Note that the view perspective in these 3D plots with the lattice bands is selected from the bottom, as it allows one to clearly see that band edges (for any band $\nu$) in the presence of SOC are internal points of the Brillouin zone. (c) Bands and gaps (shaded and white regions, respectively) in the plane of $(\lambda,\mu)$. In this panel, the edges of the band with a given index $\nu$ are defined as minimal/maximal values of $\mu_\nu$ with respect to all Bloch momentum components $k_{\mathrm{x}},k_{\mathrm{y}},k_{\mathrm{z}}$ within the Brillouin zone. The depth of the OL potential is $\mathcal{U}_0=1$ here and in all figures below.

To understand the mechanism of the formation of 3D solitons under the action of the 3D OL, it is instructive to consider, first, the linear spectrum by setting $g=0$ in Eq. (1) and searching for the Bloch eigenmodes $\mathbf{\Psi}(\mathbf{r},t)=\mathbf{u}_\nu(\mathbf{r},\mathbf{k})\exp[i\mathbf{kr}-i\mu_\nu(\mathbf{k})t]$ of the 3D OL, where $\mathbf{u}_\nu=(u_1,u_2)^{\mathrm{T}}$ is a spatially periodic constituent of the Bloch wave with periods $d_{x,y,z}$, $\nu$ is the band index ($\nu=1$ corresponds to the lowest band), $\mathbf{k}=(k_{\mathrm{x}},k_{\mathrm{y}},k_{\mathrm{z}})$ is the 3D Bloch momentum defined in the Brillouin zone (BZ) of width $\mathrm{K}_{x,y,z}=2\pi/d_{x,y,z}$, and $\mu_\nu(\mathbf{k})$ is the chemical potential (eigenvalue) of the Bloch wave that is a periodic function of the Bloch momentum $\mathbf{k}$. Typical dependencies of the eigenvalues $\mu_\nu$ for six lowest OL bands on momentum components $k_{\mathrm{y,z}}$ at $k_{\mathrm{x}}=0$ (other component sets yield similar dependencies) are plotted in Figs. 1(a) and (b) for SOC strengths $\lambda=1$ and $2$, with $\mathcal{U}_0=1$. While the bands can overlap, a sufficiently wide first finite spectral gap persists between the second and third bands for our parameters. The edges of the first finite and semi-infinite gaps can be achieved in the internal points of the BZ [see Fig. 1(b), where blue minima of the lowest band are located at intermediate values of $k_{\mathrm{y,z}}$, which is an unusual spectral feature caused by SOC]. Therefore, for sufficiently large $\lambda$, solitons bifurcating from such internal points of the spectral gaps feature spatial modulation defined by the values of $\mathbf{k}$ at these points.

With the increase of SOC strength $\lambda$ the edges of the complete gaps [white regions in Fig. 1(c)], shift to lower values of $\mu$, and bands gradually shrink. Hereafter, we focus on solitons in the semi-infinite gap and first finite gap above it, although a second finite gap opens above the critical value $\lambda\approx0.9$ too.

Lattice solitons are looked for as $\mathbf{\Psi}(\mathbf{r},t)=\psi(\mathbf{r})\exp(-i\mu t)$, where the complex spinor stationary wavefunction $\psi=(\psi_1,\psi_2)^{\mathrm{T}}$ satisfies the equation

$$\mu\psi=-\frac{1}{2}\nabla^2\psi-i\lambda(\nabla\cdot\boldsymbol{\sigma})\psi+\mathcal{U}(\mathbf{r})\psi+g(\psi^\dagger\psi)\psi, \quad (2)$$

that we solved using the modified squared operator method (MSOM) [106]. In this method the core iterative step updates the solution $\psi_n$ of Eq. (2) for profiles of stationary soliton states at the iteration $n$ to obtain solution $\psi_{n+1}$ at the next iteration, using the following scheme:

$$\psi_{n+1} = \psi_n - [\mathbf{M}^{-1}\mathbf{L}_1^{\dagger}\mathbf{M}^{-1}\mathbf{L}_0\psi - \alpha_n \langle \mathbf{H}_n, \mathbf{L}_1^{\dagger}\mathbf{M}^{-1}\mathbf{L}_0\psi \rangle \mathbf{H}_n]_{\psi=\psi_n} d\xi, \quad (3)$$

where the coefficient

$$\alpha_n = \langle \mathbf{M}\mathbf{H}_n, \mathbf{H}_n \rangle^{-1} - (\langle \mathbf{L}_1\mathbf{H}_n, \mathbf{M}^{-1}\mathbf{L}_1\mathbf{H}_n \rangle d\xi)^{-1}. \quad (4)$$

Here, $\mathbf{L}_0\psi = 0$ is the original equation (2) for stationary soliton solution, written using operator $\mathbf{L}_0$, $\mathbf{L}_1$ is the linearized operator produced by the equation for small correction $\delta\psi$ to solution $\psi$, the operator $\mathbf{M}$ is a positive-definite preconditioner (accelerator) operator (see [106] for details), and $d\xi$ is a sufficiently small step that is chosen to achieve optimal convergence. The term $\mathbf{H}_n$ is a user-selected function estimating the dominant error mode that has to be suppressed to accelerate the convergence to the exact solution. The scalar weight $\alpha_n$ is calculated to optimally suppress this mode. A common and efficient choice for this function is $\mathbf{H}_n = \psi_n - \psi_{n-1}$. The MSOM method is particularly efficient for multidimensional problems and is proven to converge to any soliton state—the ground or excited one—from a sufficiently close initial guess, provided that step $d\xi$ is sufficiently small. In this article, the grid resolution is set to $dx = dy = dz = 0.1$, the size of the integration window is set to $L_x, L_y, L_z = 40$. Upon evaluation of the action of operators $\mathbf{L}_0, \mathbf{L}_1$ on solution $\psi$ in MSOM one typically uses fast-Fourier transforms [106], hence the boundary conditions in this method are periodic. Since solitons are exponentially localized, the wavefunction $\psi$ effectively vanishes at the borders of large integration window.

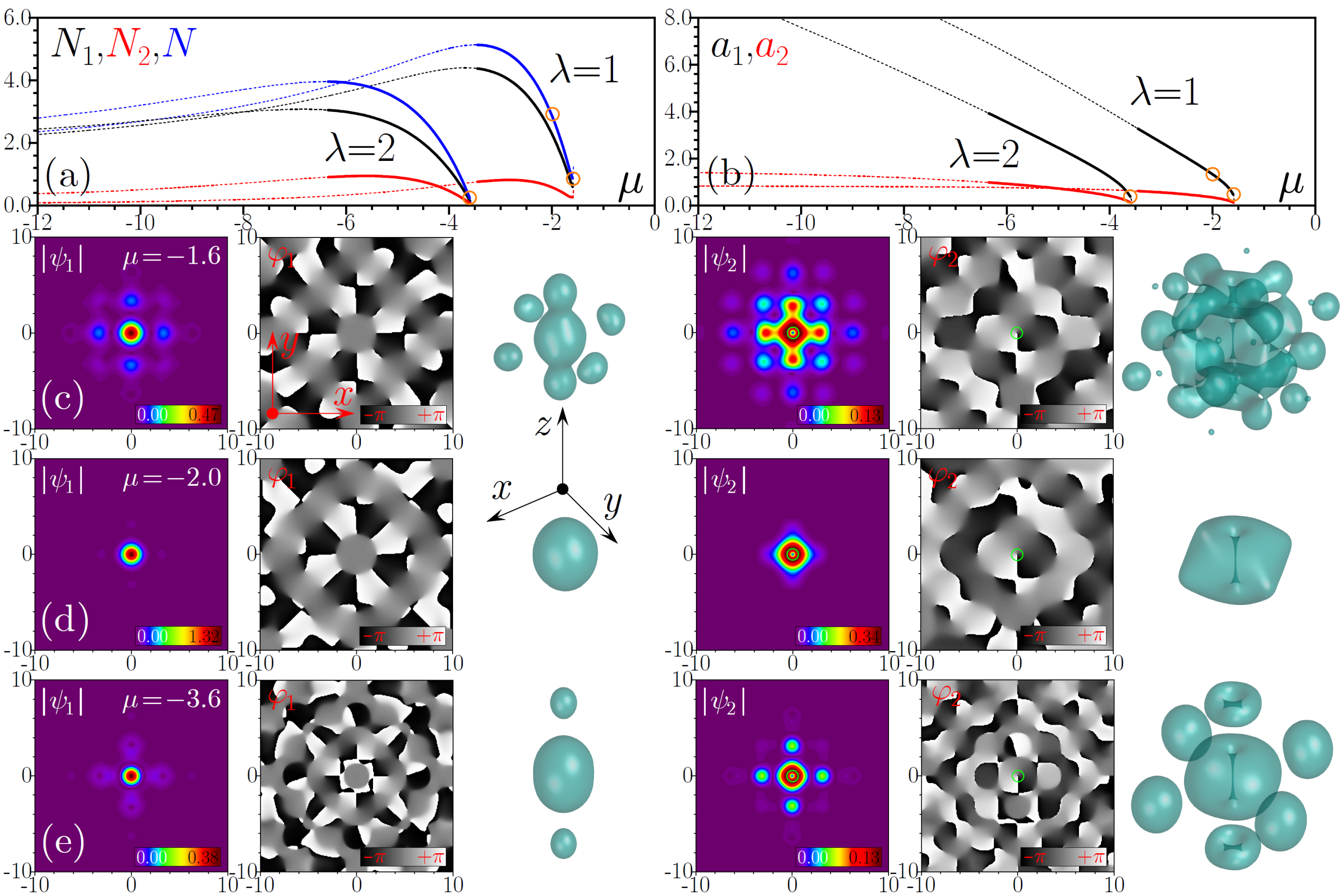

Fig. 2. Dependencies of (a) the total norm $N$ and component norms $N_{1,2}$, and (b) amplitudes $a_{1,2}$ of the components of the SV solitons on chemical potential $\mu$, for the SOC strengths $\lambda = 1$ and $2$ in the attractive binary BEC $(g = -1)$. Solid thick and dashed thin lines represent stable and unstable solitons, respectively. Orange circles correspond to the SV solitons with $\mu = -1.6$ (c) and $\mu = -2$ (d) at $\lambda = 1$ and soliton with $\mu = -3.6$ (e) at $\lambda = 2$ shown in three bottom rows. In each of these rows the absolute value and phases of both wavefunction components are plotted in the cross-section $z = 0$, along with the corresponding isosurfaces of $|\psi_{1,2}| = 0.1\max|\psi_{1,2}|$. The patterns in panels (c)-(e) are plotted in the domain of $x, y \in [-10, +10]$, while the actual integration domain is wider. Green circles in the absolute-value and phase distributions of the $\psi_2$ component indicate the location of the central vortex line in these SV states.

We start with the SV solitons (in this case the initial condition for the application of MSOM is selected in the form of a Gaussian for the first component of wavefunction $\mathbf{\Psi}$, while the second component is initially zero), representing fundamental states of the system in semi-infinite gap in the case of the self-attraction $(g = -1)$. The dependencies of the total norm of the SV solitons, $N = N_1 + N_2 \equiv \iiint (|\psi_1|^2 + |\psi_2|^2) dxdydz$ , and component norms $N_{1,2}$ on chemical potential $\mu$ are plotted in Fig. 2(a) for the SOC strengths $\lambda = 1$ and $2$ , while typical examples of the solitons (distributions of the absolute value and phase in the cross-section $z = 0$ and isosurfaces at $|\psi_{1,2}| = 0.1\max|\psi_{1,2}|$ levels) are shown in Figs. 2(c-e). In this case the vorticity $S = +1$ is carried by the weaker $\psi_2$ component (the central vortex line is evident in it in isosurface plots), while the stronger component, $\psi_1$ , has $S = 0$ . The OL-SOC interplay gives rise to multiple off-center phase singularities. The SV solitons bifurcate from the bottom edge of the first band, near which they steeply expand across the lattice, featuring multiple shape oscillations as shown in Fig. 2(c). Amplitudes $a_{1,2} = \max|\psi_{1,2}|$ of the components vanish at the gap edge and increase monotonously as $\mu$ moves away from the edge [Fig. 2(b)]. Remarkably, the OL-SOC interplay renders the dependence $N(\mu)$ nonmonotonous. Thus $N$ quickly increases with $\mu$ in a very narrow region near the top edge of the semi-infinite gap (see the dependence for $\lambda = 1$; for $\lambda = 2$ this segment is practically invisible on the scale of the plot). Adjacent to this region is a part of the family [it is shown by thick lines in Figs. 2(a),(b)] with the negative slope, $dN / d\mu < 0$ . In these branches the Vakhitov-Kolokolov (VK) stability criterion [9,107] admits stability of the SV family, because this family represents the simplest fundamental soliton states of the system. Namely, this criterion states that fundamental solitons with $dN / d\mu < 0$ may be stable, while their counterparts belonging to parts of the family with $dN / d\mu > 0$ are exponentially unstable. With the decrease of $\mu$ the solitons gradually contract [Fig. 2(d)] and, for sufficiently small values of $\mu$ on the branch with $dN / d\mu > 0$ , the solitons shrink practically to the central well of the OL potential, ceasing to be affected by the lattice (the norm approaches its value for the 3D solitons supported by SOC in the free space [86]). In this regime, SOC alone cannot stabilize high-amplitude solitons, and the corresponding branch becomes unstable, as it disobeys the VK criterion. By comparing $N(\mu)$ curves for different values of the SOC strengths $\lambda$ , one infers that stable SV solitons exist in finite intervals of the norm, that shrink with the increase of $\lambda$ , but, at the same time, the stability interval of the SV solitons conspicuously expands on the $\mu$ axis with the increase of $\lambda$ . The latter interval also expands with the increase of the OL depth $\mathcal{U}_0$ . Ratios of the component norms, $N_1 / N_2$ , and amplitudes, $a_1 / a_2$ , increase with decrease of $\mu$ , i.e., the zero-vorticity $\psi_1$ component dominates in strongly localized 3D lattice solitons.

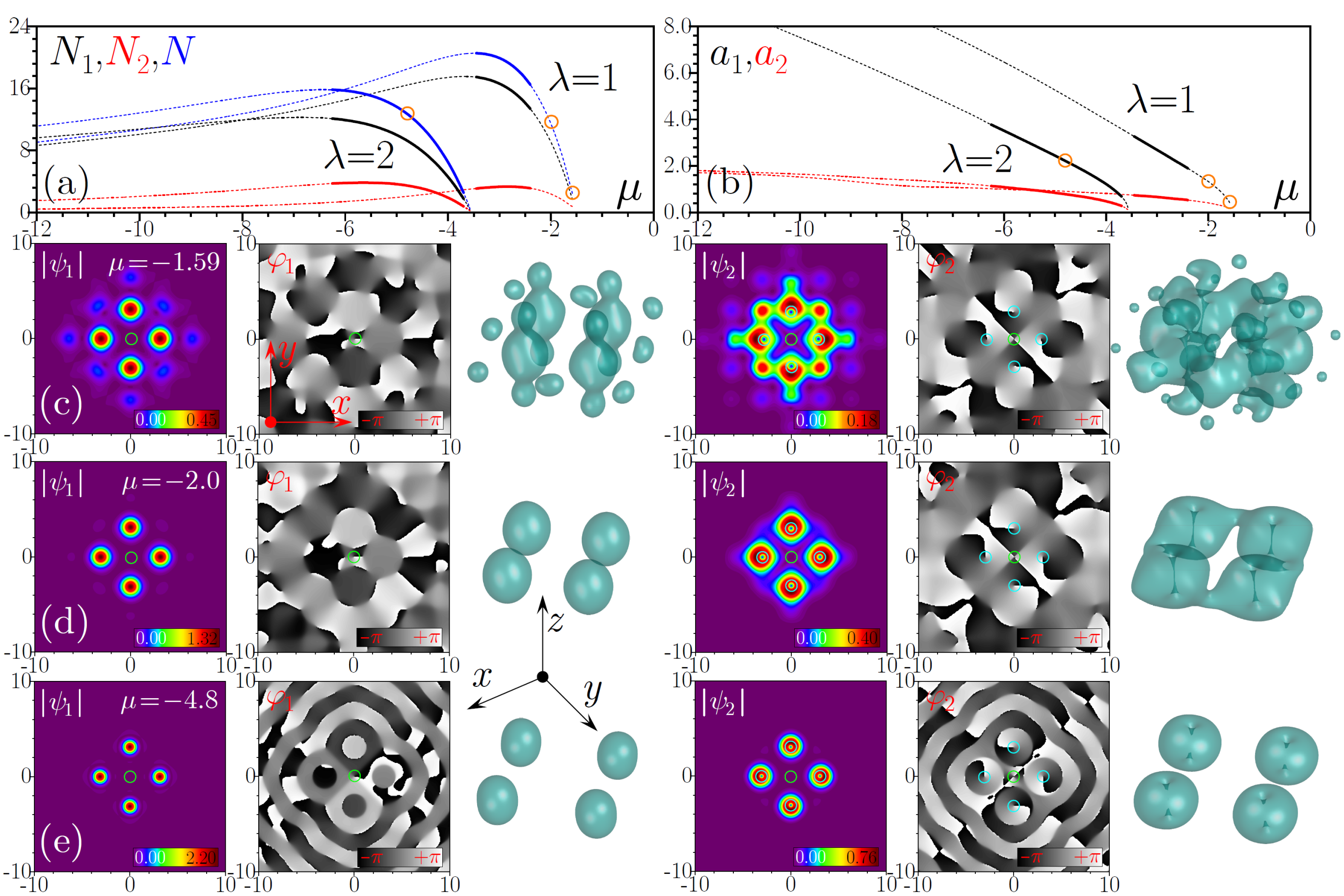

Fig. 3. The same arrangement of panels as in Fig. 2, but for the on-site full-vortex (FV) solitons in the self-attractive binary BEC $(g = -1)$ . Solid thick and dashed thin lines represent stable and unstable solitons, respectively. The absolute value and phase, as well as isosurfaces of the absolute value, are plotted for the FV solitons with $\mu = -1.59$ (c) and $\mu = -2$ (d) at $\lambda = 1$, and for $\mu = -4.8$ at $\lambda = 2$ (e), corresponding to the orange dots in (a) and (b). The patterns in panels (c)-(e) are plotted in the domain of $x, y \in [-10, +10]$ . Green and blue circles in the absolute-value and phase distributions of the $\psi_1$ and $\psi_2$ components indicate the location of central and off-center vortex lines in the FV states. Note 0that the central vortex line is not visible in the isosurface plots because the local density is strongly modulated by the lattice, taking small values close to the center of the FV states.

Comparison of phase distributions in the SV solitons at $\lambda = 2$ [Fig. 2(e)] and $\lambda = 1$ [Figs. 2(c),(d)], which are plotted in the same spatial domain, reveals attenuation of the characteristic features in the phase structure with the increase of $\lambda$ (they, in particular, determine the location of off-center vortices emerging in both components). Note that, if the spinor wavefunction $(\psi_1, \psi_2)^{\mathrm{T}}$ is a solution of Eq. (2), then $(\psi_2^*, -\psi_1^*)^{\mathrm{T}}$ is a solution too, i.e.. in addition to the SV with the vorticity-free component $\psi_1$ and vortical one $\psi_2$ carrying winding number $S = +1$, there exists the mirror-image SV with the same norm and $S = -1$ in $\psi_1$ and $S = 0$ in $\psi_2$. These conclusions, stemming from symmetries of Eq. (2), are as well valid for vortex complexes discussed below.

Another major result of the present work is that the 3D OL allows one to construct stable full-vortex solitons with nonzero vorticities in both components (recall they are complete unstable in SOC system in the free space). The initial condition for the MSOM in this case for first component was chosen in the form of a set of four Gaussians pinned to four lattice sites, with global vorticity $S = +1$ imposed on them, while the second component was initially zero. Their properties in the semi-infinite gap are summarized in Fig. 3 for the attractive interactions. The $\psi_1$ component of such solitons in any plane with fixed $z$ features four density peaks forming a rhombus residing on neighboring lattice sites with global vorticity $S = +1$ imprinted onto them [Figs. 3(c),(d)]. Under the action of SOC, the respective $\psi_2$ component is arranged as a rhombic set of four peaks with nested winding number $+1$ in each of them, but carrying also the superimposed global vorticity imposed on this component. Phase distributions are rather complicated in this case: For example, the phase singularity at the center of the $\psi_2$ component has $S = -2$ , while in $\psi_1$ component several separated singularities appear around the center. Such *stable* excited states, which are akin to on-site vortex solitons in the nonlinear Schrödinger equation with the OL potential [36,37], have not been previously reported in 3D SOC-BEC and in any other spinor system.

Isosurfaces in Figs. 3(c)-3(e) clearly show the presence of the central and four off-center vortex lines in the $\psi_2$ component, the latter ones being not parallel due to the interaction between the localized condensates. Being excited states of the system, the FV solitons do not bifurcate from the gap edge; accordingly, their $N(\mu)$ [Fig. 3(a)] and $a_{1,2}(\mu)$ [Fig. 3(b)] dependencies terminate at certain critical value of $\mu$ inside the semi-infinite gap, slightly below its top edge. At this point, the FV solitons significantly broaden and cover multiple lattice wells, but they do not expand so strongly across the entire lattice as it happens with semi-vortices. Naturally, the norm of the four-peak FV states exceeds that of the SV state approximately by a factor of $4$. The $N(\mu)$ curves are non-monotonous, and VK-stable segments with $dN/d\mu < 0$ also appear in them (recall, however, that this criterion applies only to the simplest fundamental states and does not predict full stability of excited states, such as the FV solitons). With the further decrease of $\mu$, the solitons transform into sets of four strongly localized condensates confined in respective OL wells, with vortex lines well visible in $\psi_2$ component. Actually, the variety of FV solitons can be very rich: For example, constituent condensates can have their centers pinned to the vertical, rather than horizontal plane, or arranged into fully 3D off-plane configurations.

SV and FV solitons are also supported, as gap solitons, by the system with the repulsive interatomic interactions $(g = +1)$ in a sufficiently wide first finite gap, which is clearly visible in the linear lattice spectrum displayed in Fig. 1(c). Such solitons were obtained using MSOM with the same types of inputs that were used for obtaining the vortex solitons in the case of the self-attractive BEC. SV solitons bifurcate from the bottom edge of the gap, with the amplitude increasing toward the gap's upper edge. Such solitons strongly expand across the lattice when they are placed near the lower and upper edges of the gap. A representative profile of such a state is displayed in Fig. 4(a) for chemical potential $\mu = -1$ in the middle of the finite gap, where it features the most localized form. The phase distribution in both components is more complicated in comparison to the solitons in the semi-infinite gap, supported by the attractive nonlinearity, which is a representative feature of gap solitons whose tails are usually sign-alternating ones. In Fig. 4(b) we also show an example of a full-vortex soliton, with the vorticity present in both components, while the local phase structure near the soliton's center is similar to that in the solitons belonging to the semi-infinite gap of the attractive system. We have found that, in the case of the self-repulsion, the SV solitons are stable in the first finite gap and their stability can be judged by the "anti-VK" criterion, which predicts the stability for branches with $dN/d\mu > 0$. However, the FV solitons in the first finite gap are completely unstable, and their stability properties (as of excited states) cannot be captured by this criterion, which is not a sufficient stability condition, but only a necessary one.

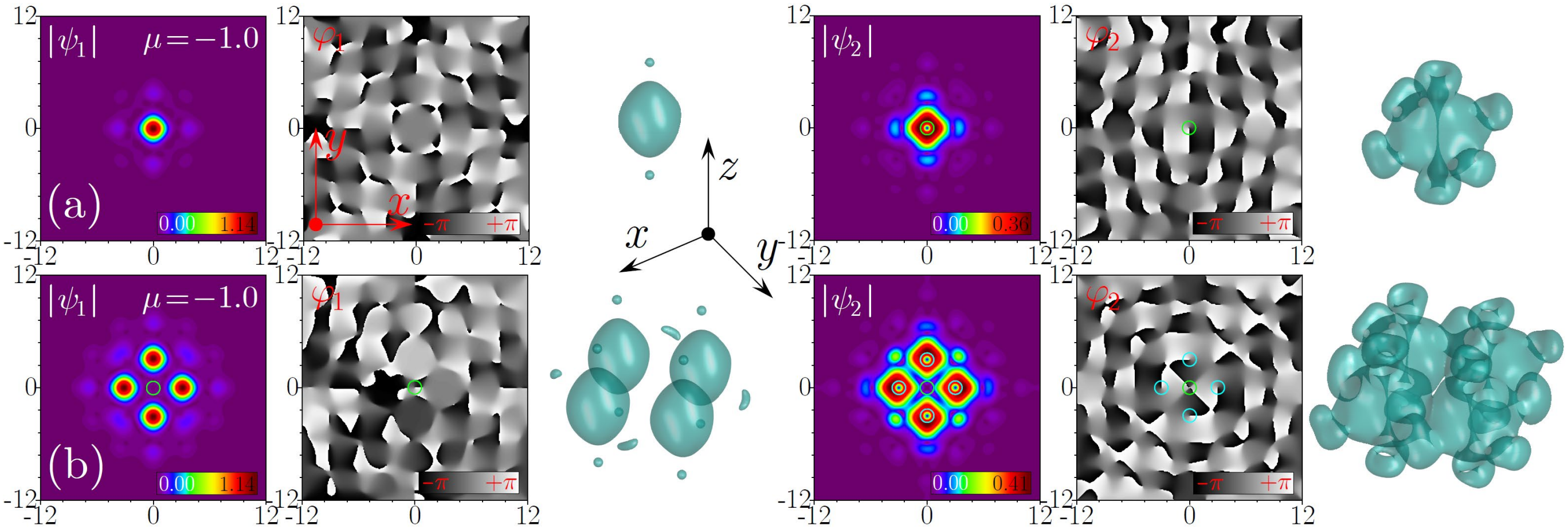

Fig. 4. The absolute value and phase of the two wavefunction components in the cross section $z = 0$ , and isosurfaces of $|\psi_{1,2}| = 0.1\max(|\psi_{1,2}|)$, which represent the gap solitons of the SV (a) and FV (b) types with $\mu = -1$ and $\lambda = 1$ in the self-repulsive BEC $(g = +1)$. The patterns are shown in the domain of $x, y \in [-12, +12]$. Green and cyan circles in the absolute-value and phase distributions indicate the location of the central vortex line in the SV states and central and off-center vortex lines in the FV ones.

Remarkably, the OL-SOC interplay allows the stabilization in the attractive system not only of SV solitons, but also of full-vortex solitons, in spite of their complicated internal structure. We checked the stability of the 3D solitons by simulating their perturbed evolution in the framework of Eq. (1) up to $t = 10^4$ with input $\boldsymbol{\Psi}(\mathbf{r}, t = 0) = \boldsymbol{\psi}(\mathbf{r})[1 + \xi(\mathbf{r})]$, where $\boldsymbol{\psi}(\mathbf{r})$ is the stationary state, and $\xi(\mathbf{r})$ represents the initial random perturbations at the 6% amplitude level (the perturbation was made different for the two components). Note that the calculated stability border is not sensitive to the amplitude of perturbations, if it remains small, but the evolution of unstable states is obviously affected by the noise amplitude. For simulations of the evolution of 3D SV and FV solitons, we used the split-step fast Fourier method with the window size $x, y, z \in [-20, +20]$, the number of grid points $400 \times 400 \times 400$ and time step $dt = 10^{-3}$. Large evolution time allows us to detect even weak instabilities of perturbed solitons. The systematic simulations confirm the full stability of the SV branches satisfying the VK criterion, $dN / d\mu < 0$, shown by thick lines in Fig. 2(a). The example of the stable evolution of a perturbed SV belonging to this branch is presented in Fig. 5(a). We plot the maximum amplitudes of both components, which remain close to their initial values, and isosurfaces of $\psi_{1,2}$, to stress that the internal structure of the soliton does not change in the course of the evolution. Broad solitons with $\mu$ close to the upper edge of the semi-infinite gap and belonging to a narrow region with $dN / d\mu > 0$ strongly broaden as the result of the development of the instability, while high-amplitude strongly localized unstable solitons with large negative values of $\mu$ suffer the collapse, see an example in Fig. 5(b), in which the dominating $\psi_1$ component collapses first.

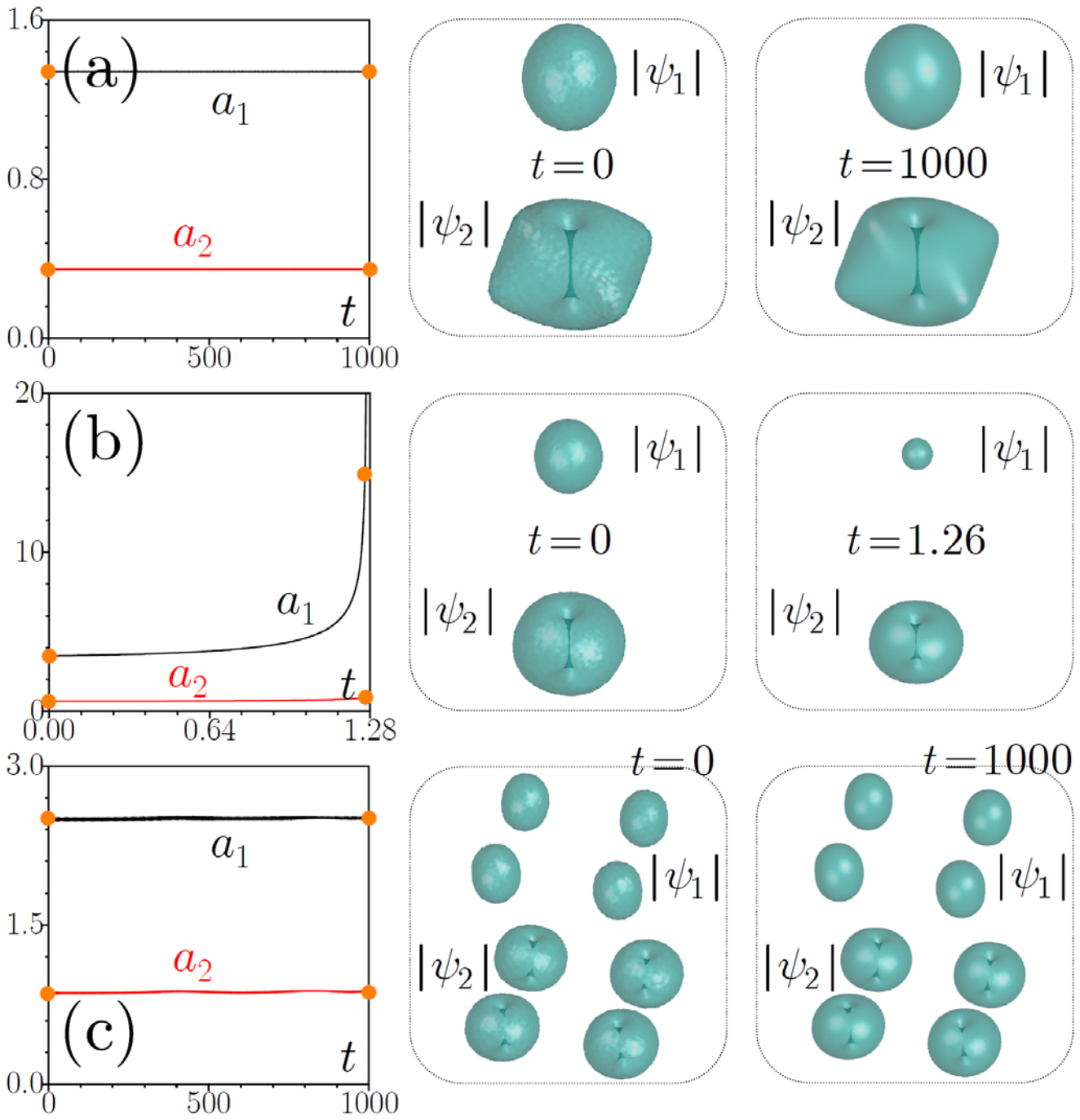

Fig. 5. Stable (a) and unstable (b) perturbed evolution of SV solitons with $\mu = -2$ and $\mu = -3.6$ at $\lambda = 1$, respectively, in the attractive system $(g = -1)$. (c) The stable evolution of the FV soliton with $\mu = -5$ at $\lambda = 2$. The left column displays amplitudes $a_{1,2} = \max|\psi_{1,2}|$ of the two components, while the central and right columns shows show isosurfaces at the $0.1\max|\psi_{1,2}|$ level at time moments corresponding to the orange dots in the left column.

Broad stability domains were also found for the FV solitons in the attractive system, see thick lines in Fig. 3(a). The stability domain occupies only a part of the VK region with $dN / d\mu < 0$, as broad full-vortex complexes with values of $\mu$ close to the gap edge are unstable and the complexes constructed of strongly localized components are unstable too. The stability interval for the FV states dramatically expands, in terms of the norm and chemical potential alike, with the increase of the SOC strength $\lambda$, stressing the importance of the OL-SOC interplay for the stabilization of the full-vortex modes. An example of the stable evolution of the perturbed FV soliton is presented in Fig. 5(c). The instability development for the low-amplitude FV solitons is accompanied by shape oscillations and broadening, while high-amplitude solitons demonstrate independent collapse of the four density peaks which form the vortices. The SV soliton in the repulsive system shown in Fig. 4(a), is also stable, while the FV soliton in Fig. 4(b) exhibits weak oscillatory instability.

In conclusion, we have produced, for the first time to our knowledge, the families of stable 3D BEC solitons supported by the interplay of SOC (spin-orbit coupling) and 3D OL (optical lattice) with both attractive and repulsive nonlinearities. The system maintains the stability of both the fundamental SV (semi-vortex) states, with vorticity present in the single component, and excited FV states, which carry vorticity in both components.

Note that using the Feshbach resonance for inducing attractive interactions in spin-orbit-coupled BEC may enhance effective losses, such as the three-body recombination. However, these loss channels may be sufficiently weak and do not affect our conclusions about stability and properties of 3D solitons under appropriate experimental conditions. Three-body recombination strength scales as the squared density of the ultra-cold gas [109], so by working with sufficiently low densities, $< 10^{12}\ \mathrm{cm}^{-3}$, one ensures that the recombination time can be kept well above the typical lifetime or typical measurement time. One can operate on the broad-resonance wing or use a background Feshbach resonance with low inelasticity, which makes it possible to sufficiently reduce the three-body recombination losses [110]. Bright solitons have been experimentally realized in self-attractive condensates of $^7$Li and $^{85}$Rb atoms, where three-body losses were manageable [111,112]. Following the first works which reported the creation of SOC systems [58], subsequent experimental results made it possible to observe attractive self-interactions in SOC condensates [113], which are central to the formation of solitons of the type discussed in our manuscript.

The analysis performed here may be naturally extended to more general configurations, including the excited states with higher vorticity, bound states of the lattice modes, structures supported by lattices with different geometries (not necessarily periodic ones), and “supervortex” states, composed of multiple compact vortices, each one carrying intrinsic vorticity $s$, arranged into ring-like configurations with an independent global vorticity $S$ imprinted onto the ring [108]. As another extension of the work, it may be interesting to develop the analysis of the vortex solitons for dipolar BEC under the action of SOC.

**Acknowledgements**
This research is funded by the National Natural Science Foundation of China (NSFC) (62205224; 11805145; 11774068); China Scholarship Council (CSC) (202006965016); Guangdong Basic and Applied Basic Research Foundation (2026A1515010001; 2023A1515010865); research project FFUU-2024-0003 of the Institute of Spectroscopy of RAS.

**Conflicts of Interest**
The authors declare no conflicts of interest.

**Data Availability Statement**
Data available on request from the authors.